# The loss of anisotropy in MgB$_2$ with Sc substitution and its relation with the critical temperature

Sabina Ruiz-Chavarria[a], Gustavo Tavizón[b] and Pablo de la Mora[a1]
[a]Departamento de Física, Facultad de Ciencias, UNAM,
Cd. Universitaria 04510, Coyoacán, D.F., México
[b]Departamento de Física y Química Teórica, Facultad de Química,
UNAM, Cd. Universitaria 04510, Coyoacán, D.F., México
[1]e-mail: delamora@servidor.unam.mx

**Abstract**
The electrical conductivity anisotropy of the $\sigma$-bands is calculated for the (Mg,Sc)B$_2$ system using a virtual crystal model. Our results reveal that anisotropy drops with relatively little scandium content (< 30%); this behaviour coincides with the lowering of $T_c$ and the reduction of the Kohn anomaly. This anisotropy loss is also found in the Al and C doped systems. In this work it is argued that the anisotropy, or *2D*-character, of the $\sigma$-bands is an important parameter for the understanding of the high $T_c$ found in MgB$_2$.

**Introduction**
Shortly after the discovery of 40-K superconductivity in MgB$_2$, the electronic nature of this system and its relation with the superconductivity was intensively studied[1,2]. MgB$_2$ is a many-band superconductor, with two bi-dimensional $\sigma$-bands and two three-dimensional $\pi$-bands at the Fermi energy. The dominant contribution to the superconductivity comes from the $\sigma$-bands[3]. When magnesium is substituted by aluminium or boron is substituted by carbon, the $\sigma$-levels start to fill up and $T_c$ diminishes and disappears when there are no more available $\sigma$-levels. De la Peña *et al*[4] showed that in the aluminium doped system $T_c$ is proportional to the number of $\sigma$-carriers; recently Kortus *et al.*[5] argued that aluminium and carbon doping can be understood in terms of band-filling and interband scattering, while de la Mora *et al*[6] found that such proportionality is clearly related to the inplane $\sigma$-electrical conductivity.

Magnesium, in MgB$_2$, has also been experimentally substituted by scandium to form the superconducting ternary system (Mg$_{1-x}$Sc$_x$)B$_2$[7,8], for *0 < x < 0.27*. Meanwhile the fully doped compound ScB$_2$ has a low $T_c$ (~ 1.5K)[9]. Electronic structure results show several different effects of Mg substitution by Sc in MgB$_2$, among these effects there are two that should be remarked. First, there are large changes in the morphology of the $\sigma$-bands, suggesting a change of dimensionality of the electronic process. Second, $E_F$ raises and comes close to the $\sigma$-band edge, but does not go above it; as a consequence the $\sigma$-band conductivity in the *a*-direction, $\sigma_a^\sigma$, reduces but does not disappear, as was the case in (Mg,Al)B$_2$ and Mg(B,C)$_2$. Calculations show that the ScB$_2$ $\sigma$-band anisotropy, $\sigma_a^\sigma/\sigma_c^\sigma = 3.9$, is much lower than for MgB$_2$ (= 43)[6]. On the other hand, in (Mg$_{1-x}$Sc$_x$)B$_2$, the experimental results show that with only *x = 0.27* then $T_c$ reduces to 6.2K. These comparative results reveal that $T_c$ drop cannot be explained by the reduction of $\sigma$-band carriers alone, as was the case for (Mg,Al)B$_2$ and Mg(B,C)$_2$ (as discussed by Kortus *et al.*[5]), but it seems that the large reduction of the $\sigma$-band anisotropy plays an important role in this system.

As suggested by the chemical composition of the system (Mg,Sc)B2, the method of supercells was initially tried for the band structure calculations, but it was found to be inappropriate since it changes the band anisotropies and also mixes the $\sigma$- and $\pi$-bands; therefore the anisotropy value is much higher than it should be, and the pure $\sigma$-band contribution can no longer be extracted. Opposite to this method, a model for MgB$_2$ was designed by replacing Mg by the (K,Ca) virtual atom with a charge of 19.4e. After this the calculations of the solid solution with Sc could be done again with the approximation of the virtual atom (virtual crystal approximation). This methodology has been selected in order to preserve the scenario in which it is possible to analyze separately the topological nature of the $\sigma$- and $\pi$-bands in one system where the $T_c$ value seems to be closely related to the $\sigma$-band anisotropy. Within the virtual crystal approximation it was found that even with little scandium substitution the $\sigma$-band anisotropy diminishes rapidly, then at $x \sim 0.3$ it levels off and continues to diminish but at a much lower rate. The rapid anisotropy diminution coincides with the experimental drop of $T_c$, revealing that the anisotropy of the $\sigma$-bands is an essential characteristic of the high $T_c$ value in MgB$_2$, in other words, our calculations show that when the $\sigma$-bands lose their two-dimensional character the compound becomes a low $T_c$ superconductor.

**Computational procedure**
The electronic structure calculations were done using the *WIEN2k* code[10], which is a Full Potential-Linearized Augmented Plane Wave (*FP-LAPW*) method based on *DFT*. The Generalized Gradient Approximation of Perdew, Burke and Ernzerhof[11] was used for the treatment of the exchange-correlation interactions. The energy threshold to separate localized and non-localized electronic states was -6 Ry. For the number of plane waves the used criterion was $R_{MT}^{min}$ (Muffin Tin radius) × $K_{max}$ (for the plane waves) = 9. The number of k-points used, for single-cell calculations, was 19×19×15 (320 in the irreducible wedge of the Brillouin zone). The magnesium muffin-tin radius=1.8$a_0$ and for boron=1.53$a_0$ ($a_0$ is the Bohr radius). The charge density criterion with a threshold of $10^{-4}$ was used for the convergence. A denser mesh of 100×100×76 (34,476 in the irreducible wedge) was used for the evaluation of the electrical conductivity. For the case of the MgScB$_2$ supercell, due to reciprocal-cell reduction, only half of the k-points were used.

**Band anisotropy**
*a) theory*
The electrical conductivity was calculated, within the relaxation time approximation and at $T = 0$, as in de la Mora *et al* and references within[6].

$$\sigma_\alpha^\beta = \frac{e^2 \tau^\beta}{\hbar \Omega_0} \int dA_\alpha \sum_i \left| v_\alpha^{i\beta}(k_F) \right| \qquad 1$$

where $\beta$ is the band index, $\tau$ is the relaxation time, $\Omega_0$ is the reciprocal-cell volume, $A_\alpha$ is the area perpendicular to the $\alpha$-direction, $v_\alpha^{i\beta}$ is the electron velocity in the $\alpha$-direction and can be calculated as the slope of the $\beta$-band (=$1/\hbar \, \partial\varepsilon^\beta/\partial k_\alpha$), $v_\alpha^{i\beta}(k_F)$ is evaluated at $E_F$, and the sum over $i$ is for all the crossings of the $\beta$-band at $E_F$. From this equation the band anisotropies can be evaluated with $\sigma_a^\beta/\sigma_c^\beta$.

The band anisotropy can approximately be analyzed with the expression[6]:

$$\frac{\sigma_a^\beta}{\sigma_c^\beta} \approx \frac{v_a^2}{v_c^2} \approx \frac{A_a^2}{A_c^2} \qquad 2$$

The relation between conductivities, $\sigma_a^\beta/\sigma_c^\beta$, and areas, $A_a^2/A_c^2$, was found to be correct within 17% for the (Mg,Al)B$_2$ system[6].

*b) discussion*

The objective of this paper is to study theoretically the effect on the $\sigma$-band anisotropy, $\sigma_a^\sigma/\sigma_c^\sigma$ ($\sigma_a^\beta$ is the conductivity of the $\beta$-band in the $\alpha$-direction and should not be confused with $\sigma$, which is used to refer to bands), of the scandium substitution on MgB$_2$, and how this anisotropy can be related to the experimental reduction of $T_c$.

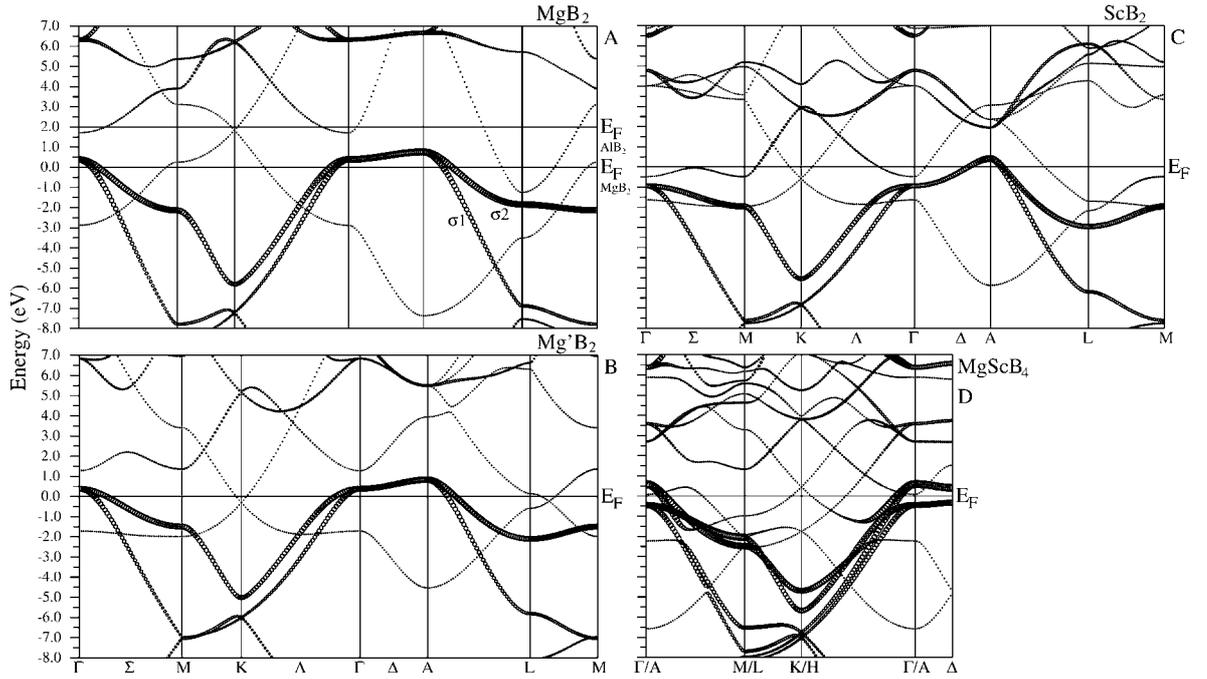

**Figure 1. Band structure of a)MgB$_2$, the thick lines correspond to the $\sigma$-bands. AlB$_2$ has a quite similar band structure, but with $E_F$ at approx. 2eV above, b) Mg'B$_2$ adjusted model (see text), the $\sigma$-bands are quite similar to MgB$_2$, c) ScB$_2$, $E_F$ is higher than in MgB$_2$, but not above the $\sigma$-bands, also the $\sigma$-band slope in the $c$-direction ($\Gamma$-A, L-M) is considerably higher, and d) MgScB$_4$ calculated with the supercell model, the symmetry labels correspond to the single cell labels so that it can be compared with other band structures. Note the splitting that appears in the $\sigma$-bands at $\Delta$.**

The band structure of MgB$_2$ and ScB$_2$ can be seen in figure 1 (*a*) and (*c*), the thick lines near the Fermi energy ($E_F$) are the $\sigma$-bands; $\sigma 1$ being the lower one and $\sigma 2$ the upper one. There are two other bands, the $\pi$-bands, with thin lines.

Observing the band structure of MgB$_2$ it can be seen that in the *a-b* plane ($\Gamma$-M-K-$\Gamma$ and A-L) the $\sigma$-bands have a large slope, in contrast, in the *c*-direction ($\Gamma$-A and L-M) they have a small slope, therefore these bands can be expected, from expression 2, to be highly anisotropic, the calculated value for $\sigma_a^\sigma/\sigma_c^\sigma$, using equation 1, was 43.

AlB$_2$ has a very similar band structure to MgB$_2$. Since aluminium [Ne]3s$^2$3p$^1$ contributes with one electron more than magnesium [Ne]3s$^2$ then $E_F$ is about 2eV higher, which is above the $\sigma$-bands, and these bands no longer contribute to the conductivity; AlB$_2$ is not a superconductor, which is congruent with the fact that the σ-bands are the responsible of the high $T_c$ in MgB$_2$. The most prominent difference, besides the $E_F$ shift, is a lowering of the π-bands relative to the σ-bands. Due to the similarity of the band structure of these compounds, the rigid band approximation can be applied very well for the computation of $\sigma_a^\sigma$ in the (Mg,Al)B$_2$ system[6].

Scandium [Ar]4s$^2$3d$^1$ in ScB$_2$ also adds one electron more than magnesium, but in contrast to the *3p*-aluminium orbitals, the *3d*-scandium orbitals have a different symmetry and when they hybridize with the boron σ-orbitals, they reduce the *2D*-character of the latter. The corresponding electronic structure is quite different to AlB$_2$ (figure 1c); $E_F$ is still within the σ-bands but, in comparison with MgB$_2$, it has a much larger slope in the *c*-direction and smaller in the *a-b* direction; see for example *σ2* in the Γ-M section. Correspondingly, the respective σ-band anisotropy is much lower: 3.9.

For the (Mg,Al)B$_2$ and Mg(B,C)$_2$ systems the $T_c$ reduction with doping has been explained as a consequence of the reduction of the σ-band carriers[4-6]. In the (Mg,Sc)B$_2$ system the drop of $T_c$ cannot be completely related to the reduction of σ-band carriers since with 27% of scandium substitution the system has a low $T_c$ (~ 6K)[7,8], and even in ScB$_2$ the system has quite a bit of σ-band carriers ($\sigma_a^\sigma$(ScB$_2$) = 0.12$\sigma_a^\sigma$(MgB$_2$)). On the other hand, anisotropy is reduced from 43 in MgB$_2$ to 3.9 in ScB$_2$. Therefore the $T_c$ reduction cannot be ascribed to the σ-band carriers reduction but it should be associated to the loss of σ-band anisotropy. The *2D* character of the σ-bands should be considered as an important feature of the high $T_c$ in MgB$_2$.

**Model**

The (Mg$_{1-x}$Sc$_x$)B$_2$ solid solution has been obtained experimentally in the AlB$_2$ structure for $x$ = [0-0.02] and [0.12-0.27][8]. The electronic structure calculation for this solid solution presents several difficulties. One method that could be used is based in constructing supercells; it consists in repeating the unit cell *n* times, replacing magnesium by scandium in *m* of them. We found three problems with this method;

A.   When the unit cell is repeated in the *a-b* plane then the σ- and π-bands are mixed and the properties of the σ-bands cannot be easily calculated, this is specially relevant since the σ-bands are the responsible for the high $T_c$ in MgB$_2$[3-5].

B.   In the supercell model magnesium and scandium would have a crystalline order, that is, they are not disordered as in the real compound. As a consequence, certain directions may be favoured, for example, if the single unit-cell is repeated twice in the *c*-direction to construct the supercell, one unit-cell with Mg and the other with Sc, then the crystal would have alternating layers of magnesium and scandium. Magnesium and scandium produce a different electrical potential in the crystal, therefore the bidimensionality is enhanced; the σ-bands, that are already quite 2D, are split in Δ. This splitting reduces the slope of the bands in the c-direction and as a consequence $\sigma_c^\sigma$ is reduced. In the π-bands, which are 3D, the splitting in Δ (at -4.7eV) is minimal.

An extreme example would be a stack of 3 layers of MgB$_2$ followed by 3 layers of AlB$_2$, in this case AlB$_2$ does not have any σ-bands at $E_F$ and $\sigma_c^\sigma$ would be zero, but $\sigma_a^\sigma$ would be different from zero. This would be seen in the band structure as a splitting in Δ and the bands in the *c*-direction would be completely flat.

C. It has the restriction that *x* should be a fraction: *m/n*. To have a representative sampling for *x*, then *n* has to be large, which implies prohibitive computational costs.

An alternative method to surpass these problems is the virtual crystal approximation (VCA), in which the charge assigned to an atom is fractional instead of integer; this virtual atom has the mixed properties of the two contiguous atoms. This can be used for the $(Mg_{1-x}Al_x)B_2$ system[4]. In the VCA method $Mg_{1-x}Al_x$ is replaced by a virtual atom M with a charge of *12+x* electrons, resulting in the new system $M^{12+x}B_2$. This method can only be applied when the solid solution is between two contiguous atoms such as Mg and Al, therefore it cannot be applied directly to the $(Mg_{1-x}Sc_x)B_2$ system.

To overcome these limitations it was thought that $CaB_2$ can be used instead of $MgB_2$ as the initial system, since magnesium and calcium have the same valence, this is further justified since in this compound Mg is almost fully ionized (~ $Mg^{+2}$)[12]. With this replacement then the system $(Ca,Sc)B_2$ would be used instead of $(Mg,Sc)B_2$.

Calculations of $CaB_2$ showed that the σ-bands are not that similar to those of the $MgB_2$ compound, they seem to be between those of $MgB_2$ and $ScB_2$, in fact the slope in the *c*-direction has increased. For this reason it was though that instead of using calcium, maybe a virtual $K_{1-x}Ca_x$ atom could better replace magnesium. Therefore, to simulate $MgB_2$, the system $(K_{1-x}Ca_x)B_2$ was studied with the VCA; it was found that with $x = 0.4$ the resulting band structure is very similar to that of $MgB_2$. Further refinements were done on the cell parameters to adjust the bandwidths, both in the plane and in the *c*-directions. The adjusted cell parameters were: a = 3.0108Å and c = 3.9742Å (in $MgB_2$ a = 3.0864Å and c = 3.5215Å). In the adjusted cell $E_F$ was found to be lower that in $MgB_2$, therefore it was increased by $\Delta E_F = 0.04Ry$ (= 0.54eV).

The band structure of the fully adjusted cell is shown in figure 1b. It can be seen that σ-bands are quite similar to those of $MgB_2$, figure 1a. On the other hand, the π-bands are quite different, but since what is analyzed here are the σ-bands, then this difference should not influence the study of $\sigma_\alpha^\sigma$, since these bands at $E_F$ are quite separated. Our calculations show that the σ- and π-Fermi surfaces (FS) are spatially well separated, the σ-FS are around the Γ-A axis (the distance of the σ-FS to this axis is less than 0.44 of the Γ-K distance), while the π-FS are near the vertical sides of the cell (the distance to the Γ-A axis is more than 0.80 of the Γ-K distance).

Since the π-orbitals are far from the σ-orbitals, then in terms of the σ-conductivity, $\sigma_\alpha^\sigma$, they act only as recipient of electrons. This effect can be seen, but in a diminished way, in the $(Mg_{1-x}Al_x)B_2$ system[6] where the effect of the π-orbitals was to shift upwards the σ-bands when *x* increased, that is, the π-orbitals start to fill faster than the σ-orbitals and as a consequence $\sigma_\alpha^\sigma$ took longer to vanish.

With this adjustment $(K_{1-x}Ca_x)B_2$, which will be referred as $Mg'B_2$, can now be used as $MgB_2$, and the $(Mg'_{1-x}Sc_x)B_2$ solid solution can be again calculated with VCA. The cell parameters and $\Delta E_F$ for the intermediate values of $x$ were calculated as a linear interpolation between the Mg'B$_2$ and ScB$_2$ values. Then, in solid solution (Mg'$_{1-x}$Sc$_x$)B$_2$, $x$ took values in intervals of *1/8* from *0* to *1*.

Within this framework, the results of the $\sigma_a^\sigma$ calculations may not be very precise, but we consider that the general trends of the $\sigma$-band behaviour, as a consequence of the scandium doping, should be correct.

**Results and discussion**

Observing the $\sigma$-bands in figure 1, it can be seen that $\sigma 2$ in MgB$_2$ (figure 1a) has a positive slope in $\Gamma\rightarrow A$ (above $E_F$) and in M$\rightarrow$L (below $E_F$, both in the *c*-direction). On the other hand, Mg'B$_2$ (figure 1b) the slope at M$\rightarrow$L is negative, therefore at $E_F$ the band slope (and also $\sigma_c^{\sigma 2}$) is very small, this is reflected in the corresponding FS, which is an almost straight tube and the band anisotropy obtained from expression 2 is very large[6]. On the other hand, $\sigma 1$ has the same slope in both compounds and it should reproduce the correct behaviour of the anisotropy of the conductivity in Mg'B$_2$.

The $\sigma$-band conductivities for the model compound, (Mg'$_{1-x}$Sc$_x$)B$_2$, are shown in figure 2. $\sigma_a^\sigma$ and $\sigma_c^\sigma$ are shown in figure 2a, while the anisotropy, $\sigma_a^\sigma/\sigma_c^\sigma$, is shown in figure 2b.

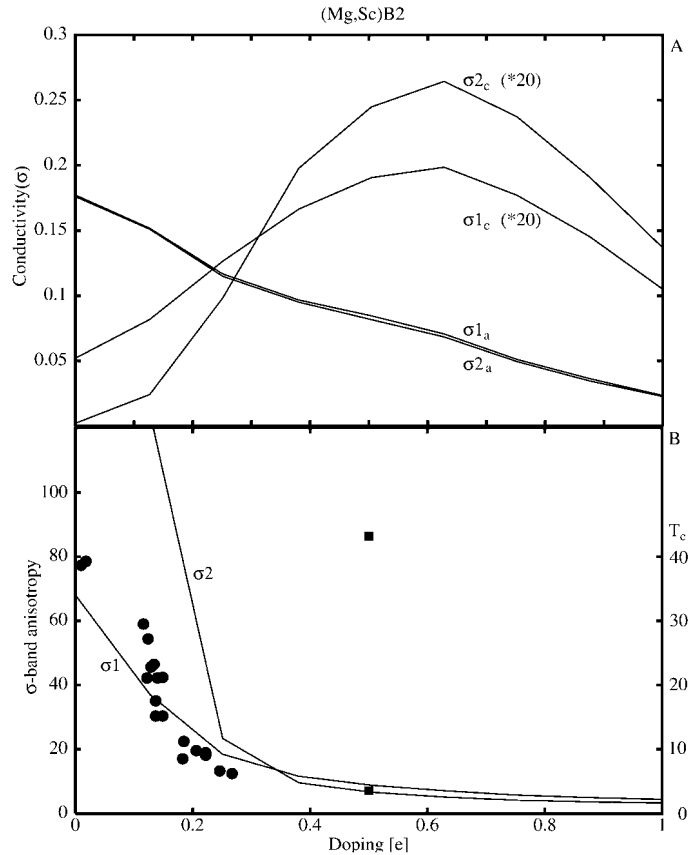

**Figure 2. a)** $\sigma$-**band conductivities (**$\sigma 1$**,** $\sigma 2$**);** $\sigma_a^\sigma$ **in the** *a-b* **plane and** $\sigma_c^\sigma$ **in the** *c*-**direction (the** *c*-**component has been multiplied by 20), b)** $\sigma$-**band anisotropies, the squares show the anisotropy of the MgScB$_4$ supercell, see text; the circles are the experimental values of T$_c$[8].**

As shown in figure 2a, the inplane conductivity, $\sigma_a^\sigma$, decreases continuously. On the other hand, the value of the *c*-direction conductivity, $\sigma_c^\sigma$, has a large increase at the beginning, then it decreases. For the $\sigma2$-conductivity, $\sigma_c^{\sigma2} \approx 0$ at $x \approx 0$, this is a consequence of the $\sigma2$-FS being an almost straight tube.

For $x = 0$ (Mg'B$_2$) $\sigma1$ has an anisotropy of *67*, which can be regarded in this model, as close to the MgB$_2$ value of *43*. For $\sigma2$, due to the low value of $\sigma_c^{\sigma2}$, the anisotropy is much higher, *1500*, then it lowers quite rapidly and for $x = 0.125$ it becomes *120*. For low scandium content ($0 < x < 0.3$) both $\sigma$-band anisotropies drop rapidly to a low value and, at $x \approx 0.3$ they have the same value. After this interception they continue to drop, but at a much lower rate; in this range ($0.4 < x < 1$) they have almost the same value ($\sigma1 \approx \sigma2 \times 1.3$). The drop of $T_c$ is in the range of low $x$, where there is a large anisotropy decrease ($0 < x < 0.3$). In this range the inplane $\sigma$-conductivity has a moderate reduction of 35%. Therefore, the fall in $T_c$ cannot be associated to the $\sigma_a^\sigma$ diminution (which is related to the band-filling) as was the case of aluminium and carbon doped MgB$_2$[4-6]. The VCA treatment of the (Mg'$_{1-x}$Sc$_x$)B$_2$ system enhanced the anisotropy at $x \approx 0$ (*67* in Mg'B$_2$ compared to the *43* in MgB$_2$), therefore in a more realistic model the value of the anisotropy at $x \approx 0.3$ should be lower. The anisotropy value, $\sigma_a^\sigma/\sigma_c^\sigma$, at $x = 0.27$ is $\approx 17$ (figure 2b), which, if adjusted, could be as low as *11*.

To further justify the (Mg'$_{1-x}$Sc$_x$)B$_2$ model, a supercell was made by piling MgB$_2$ and ScB$_2$ in the *c*-direction, for the *a* parameter the average value of the two cells was used, while for *c* the sum was used. Except for the substitutional order this cell would correspond to (Mg$_{0.5}$Sc$_{0.5}$)B$_2$. This compound consists of alternating ...B-Mg-B-Sc-B... layers, since scandium contributes less to the $\sigma$-conductivity then, as explained above, the material should have a smaller $\sigma$-conductivity in the *c*-direction and the anisotropy should be higher than that of a Mg-Sc-disordered system.

The reciprocal cell of this (MgSc)B$_4$-double cell is folded in half, the Γ-M-K and A-L-H planes of the single cell now fall into the Γ-M-K plane of the double cell, therefore the bands of both planes of the single cell can be seen in the Γ-M-K plane, and the bands in the *c*-direction are cut in half and folded. The band structure that is shown in figure 1d; the Γ-M and A-L bands are now in Γ-M, the Γ-A bands are folded, see for example the $\pi$-bands at *-0.4eV* at Δ. In contrast the $\sigma$-bands at $E_F$ are split, due, as mentioned above, to the Mg-Sc-layered arrangement.

The value of the $\sigma$-band anisotropy for the (MgSc)B$_4$ supercell is *86*, and is shown in figure 2 by the upper square. This large value is due to the large splitting in the $\sigma$-bands, seen in Δ of figure 1d**.** This splitting lowers the slope quite considerably in the *c*-direction. In the real material magnesium and scandium should be randomly ordered and there should be no splitting, as is the case for the $\pi$-bands. The $\sigma$-bands cross Γ at *0.5963eV* and *-0.4448eV* and cross Δ at *0.4054eV* and *-0.3358eV*, therefore the average slope would be *0.1500eV/δ* where $\delta$ is the Γ-A distance. If there was no splitting at Δ, then the slope would increase to *0.5202eV/δ*, that is, it would increase *3.47* times, and the conductivity would increase, using expression 2, by $3.47^2 = 12.03$ times. Therefore the anisotropy would lower from *86* to the value of *7.2*. This is shown by the lower square in figure 2. This value fits the results of the (Mg'$_{1-x}$Sc$_x$)B$_2$ model for $x = ½$ surprisingly well. Although this value cannot be taken as very precise, it shows that the VCA model is capable of reproducing some quantitative aspects of the $(Mg_{1-x}Sc_x)B_2$ system.

As an additional test to monitor the supercell methodology, a supercell was made by repeating the $MgB_2$ cell six times, three cells on the *a-b* plane and twice in the *c*-direction. Then magnesium is replaced by scandium in one of the cells; the electronic structure of this cell shows that the $\sigma$- and $\pi$-bands get strongly mixed and can no longer be regarded as independent, these bands have both the $\sigma$- and the $\pi$-character. The supercell method cannot be used to calculate $\sigma$-band anisotropies of $(Mg_{1-x}Sc_x)B_2$ for values of $x = 1/6$, 1/3, etc.; the VCA model with the virtual magnesium Mg' seems to be the simplest suitable alternative.

Agrestini *et al*[7,8] measured $T_c$ as function of scandium content in $(Mg_{1-x}Sc_x)B_2$; the values of the second paper[8] are reproduced in figure 2b (the absence of values in the 0.02 and 0.12 range, is due to a miscibility gap). This coincides with the argument of Agrestini *et al.* that the reduction of $T_c$ can be associated to a *2D/3D* electronic transition, although they call it an electronic topological transition (ETT). They associated this ETT to the point where the $\sigma$-Fermi surface tubes are split into elongated droplets and corresponds to the point when the $E_F$ crosses the bands at the $\Gamma$-point[13] (see for example the discussion in; de la Peña *et al*[4]). They expected that the scandium concentration at the ETT to be *x = 0.12*. They also find a reduction of the Kohn anomaly going from $MgB_2$ to $(Mg_{0.8}Sc_{0.2})B_2$[8]; according to them 'this can be understood if the Sc substitution has driven the chemical potential through the ETT, where the $\sigma$-FS has a *3D* topology with a reduced Kohn anomaly'. It should be noted that for $ScB_2$ there is a further reduction of the Kohn anomaly, in other words, for *x = 0.12* the Kohn anomaly has an intermediate value between those of $MgB_2$ and $ScB_2$.

A reliable estimation of the dimensionality could result from the band anisotropy relation, $\sigma_a^\sigma/\sigma_c^\sigma$, which can be estimated from expression 2 by observing the areas span from the *a*- or *c*-directions. When the Fermi surfaces of $MgB_2$, corresponding to the $\sigma$-bands, are seen from the plane they are like tubes, but from the *c*-direction they are thin rings. At the ETT the anisotropy should be already quite small; therefore the material would have a *3D*-character. At *x = 0* the $\sigma$-FS, observed from the *c*-direction, a small area is observed and the anisotropy is large; when the tubes become droplets, the ring observed from above, now becomes a solid circle and the area is no longer small, that is, $\sigma_c^\sigma$ is now large and the anisotropy becomes small. Therefore there is an anisotropy reduction in the *0 < x < 0.35* range. According to our calculations the ETT is not at *x = 0.12* but at *x = 0.35*. It should be noted that this anisotropy change although it is large, it is smooth.

At this ETT we do not find any $\sigma$-band anisotropy jump, not even a sharp change of slope (this also applies to the case of aluminium substitution, see; de la Mora *et al*[6]). This value does represent the point where the $\sigma$-band anisotropy curves have a smooth change of slope from a large one (*x < 0.35*) to a low one (*x > 0.35*). In this framework *x = 0.35* represents the composition value in which the $\sigma$-bands have reached a *3D* character. According to our methodology, for $(Mg_{1-x}Sc_x)B_2$, a smooth and continuous *2D* to *3D* electronic dimensional transition (EDT) of the $\sigma$-bands is observed, but it is not related to the $\sigma$-Fermi surfaces topological transformation (from tubes to droplets).

According to our results on electronic structure of (Mg,Sc)B$_2$ in the modality of the VCA, the EDT transition should be allocated in the region of *0.1 < x < 0.3*, which is also the region when there is a large reduction of $T_c$.

For *x = 0.2* the $\sigma 1$-band anisotropy is *~ 15-23*, that could be seen as an intermediate *2D-3D* value. Our studies on band structure of (Mg,Sc)B$_2$ based on the VCA can be used to re-interpret the results of Agrestini *et al*[8] for the Kohn anomaly: the $\sigma$-bands for MgB$_2$ are in the *2D* regime; on the other extreme that of ScB$_2$ are in the *3D* regime. In this scenario (Mg$_{0.8}$Sc$_{0.2}$)B$_2$ has an intermediate *2D-3D* anisotropy value, and this could be the reason for the intermediate value of the Kohn anomaly for this specific composition. In this way both, the Kohn anomaly value and the $\sigma 1$-band anisotropy cannot be assigned to a definite *2D* or *3D* regime.

Although $T_c$ reduction in Al and C doping has been attributed to band-filling and interband scattering[4-6], anisotropy reduction must also play an essential role in these systems. With electronic structure calculations on (Mg,Al)B$_2$ and Mg(B,C)$_2$ within the VCA, we can also show that anisotropies in these systems drop to values comparable to those of the (Mg,Sc)B$_2$ system, at the doping values where the $\sigma$-band becomes full, this doping corresponds to the lowest $T_c$ in those systems. For the calculations the cell-parameter values of de la Peña *et al*[4] for (Mg,Al)B$_2$ and those of Kazakov *et al*[14] for Mg(B,C)$_2$) were used; these anisotropy values are shown in table 1.

| System | $\sigma_a^{\sigma}/\sigma_c^{\sigma}$ |
|---|---|
| (Mg'$_{0.73}$Sc$_{0.27}$)B$_2$ | 11-17 |
| (Mg$_{0.5}$Al$_{0.5}$)B$_2$ | 7.5 |
| Mg(B$_{0.85}$C$_{0.15}$)$_2$ | 13 |

Table 1. Anisotropy values for the different systems at concentrations where $T_c$ has reached low values.

The high $T_c$ in MgB$_2$ is due to both the $\sigma$- and $\pi$- contributions, being the former one the strongest; scandium substitution, which mixes the $\sigma$- and $\pi$- bands (see figure 1d), would tend to lower $T_c$, without this mixing $T_c$ would probably drop at a higher scandium content.

For the aluminium and carbon-doped systems the reduction of $T_c$ can be explained by band-filling, but in the case of scandium doping the band-filling is not complete, in fact, it is quite limited ~ 35%. On the other hand, in the three systems the reduction of $T_c$ is accompanied by the reduction of the anisotropy, that is, a *2D→3D* change. Therefore the *2D* character or the large $\sigma$-band anisotropy should be considered as responsible for the high $T_c$ found in MgB$_2$.

**Conclusions**
The model presented here constitutes a phenomenological description based on the band anisotropy of the MgB$_2$ system. The VCA treatment, with the possibility of having independent contributions to the conductivity from the $\sigma$- and the $\pi$-bands, gives a plausible explanation of the high $T_c$ behaviour in the (Mg,Sc)B$_2$ system.

In conclusion; a theoretical model to study the (Mg$_{1-x}$Sc$_x$)B$_2$ system is presented. The model, based on the virtual crystal approximation, is able to calculate the anisotropy of the electrical conductivity of the $\sigma$-bands. It is also shown that the calculation of this anisotropy is not possible with the method of supercells. $T_c$ and the Kohn anomaly follow a quite similar tendency as the conductivity-anisotropy of the $\sigma$-bands. The loss of anisotropy is also present in the aluminium and carbon doped systems. The $\sigma$-band anisotropy, or in other words, its *2D* character, emerges as an essential factor in the superconductivity of $MgB_2$.

**Acknowledgements**

This work was done with support from DGAPA-UNAM under project PAPIIT IN 112005.

**References**
1    J. Kortus, I.I. Mazin, K.D. Belashchenko, V.P. Antropov, L.L. Boyer, Phys. Rev. Lett. **86**, 4656 (2001).
2    A comprehensible review of the first two years' research on magnesium diboride can be found in March 2003 special issue of Physica C, **385** (1-2).
3    P.C. Canfield and G.W. Crabtree, Physics Today, March 34 (2003).
4    O. de la Peña, A. Aguayo, R. de Coss, Phys. Rev. B **66**, 12511 (2002).
5    J. Kortus, O.V. Dolgov, R.K. Kremer and A.A. Golubov, Phys. Rev. Lett. **94**, 27002 (2005).
6    P. de la Mora, M. Castro, G. Tavizon, J. Phys.: Condens. Matter **17**, 965-978 (2004).
7    S. Agrestini, C. Metallo, M. Filippi, C. Sanipoli, S. De Nigri, M. Giovannini, A. Saccone, A. Latini, and A. Bianconi, Journ. of Phys. and Chem. of Solids **65**, 1479 (2004).
8    S. Agrestini, C. Metallo, M. Filippi, L. Simonelli, G. Campi, C. Sanipoli, E. Liarokapis, S. De Nigri, M. Giovannini, A. Saccone, A. Latini, and A. Bianconi, Phys. Rev. B **70**, 134514 (2004).
9    G.V. Samsonov and I.M. Vinitsky, *Refractory Compounds* (in Russian) (Metallurgija, Moskva, 1976).
10   P. Blaha, K. Schwarz, G.K.H. Madsen, D. Kvasnicka, and J. Luitz, *WIEN2k, An Augmented Plane Wave + Local Orbitals Program for Calculating Crystal Properties* Karlheinz Schwarz, Techn. Universität Wien, Austria, 2001 (ISBN 3-9501031-1-2).
11   J.P. Perdew, S. Burke, K. Ernzerhof, Phys. Rev. Lett. **77**, 3865 (1996).
12   P. de la Mora, M. Castro, G. Tavizon, J. Solid State Chem. **169**, 168 (2002).
13   A. Bianconi, S. Agrestini, D. Di Castro, G. Campi, G. Zangari, N.L. Saini, A. Saccone, S. De Nigri, M. Giovannini, G. Profeta, A. Continenza, G. Satta, S. Massidda, A. Cassetta, A. Pifferi, and M. Colapietro, Phys. Rev. B **65**, 174515 (2002).
14   S.M. Kazakov, R. Puzniak, K. Rogacki, A.V. Mironov, N.D. Zhigadlo, J. Jun, Ch. Soltmann, B. Batlogg, and J. Karpinski, Phys. Rev. B **71**, 24533 (2005).